\documentclass[journal]{IEEEtran}
\usepackage{cite}
\usepackage{amsmath,amssymb,amsfonts}
\usepackage{algorithmic}
\usepackage{graphicx}
\usepackage{textcomp}
\usepackage{xcolor}
\usepackage{hyperref}
\usepackage{flushend}

\def\BibTeX{{\rm B\kern-.05em{\sc i\kern-.025em b}\kern-.08em
    T\kern-.1667em\lower.7ex\hbox{E}\kern-.125emX}}
\usepackage{svg}
\renewcommand{\thesection}{\textbf{\Roman{section}}}

\begin{document}

\title{Deep-Cryogenic Modeling of 22-nm FDSOI MOSFETs based on BSIM-IMG}

\author{

Debargha Dutta, Kerim Ture, Fabio Olivieri, Alberto Gomez-Saiz, and Grayson M. Noah

\fbox{%
  \parbox{\textwidth}{%
    This work has been submitted to the IEEE for possible publication. 
    Copyright may be transferred without notice, after which this version 
    may no longer be accessible.
  }%
}

\thanks{This work was supported by the Industrial Strategy Challenge Fund (ISCF)
Project Altnaharra (Innovate U.K.) under Grant 10006186. The work of A. Gomez-Saiz was supported by an Industrial Fellowship from the Royal Commission for the Exhibition of 1851.}
\thanks{D. Dutta, K. Ture, F. Olivieri, and G. M. Noah are with Quantum Motion, London N7 9HJ, United Kingdom (e-mails: debargha@quantummotion.tech, grayson@quantummotion.tech).}
\thanks{A. Gomez-Saiz is with Quantum Motion, London N7 9HJ, United Kingdom, and also with the Department of Electrical and Electronic Engineering, Imperial College London, London SW7 2AZ, United Kingdom.}

}

\maketitle

\begin{abstract}
We present a modeling approach based on the BSIM-IMG compact model to capture the deep-cryogenic behavior of MOSFET devices in a 22-nm FDSOI technology. The modeling flow is based on DC measurements to extract static parameters including variability and RF measurements to extract dynamic parameters. Modifications to the mobility equations are introduced to enable the modeling of intersubband scattering effect. The extracted models are used to enable deep-cryogenic simulations of a digital-to-analog converter (DAC), showing close agreement with measurement results.
\end{abstract}


\section{\textbf{Introduction}} \label{Intro}
In the last decade there has been a growing demand for electronics to operate at deep-cryogenic temperatures ($\leq10$~K) driven by research areas such as quantum computing \cite{10.1145/3061639.3072948} and high-energy physics \cite{9467069}. In particular, CMOS technology operating in the deep-cryogenic regime (cryo-CMOS) has been identified as a promising candidate to address the quantum-classical interface I/O bottleneck present in large-scale quantum computing systems \cite{gonzalez2021scaling}.
While foundries provide accurate device models as part of their Process Design Kits (PDKs), these models generally do not cover the deep-cryogenic temperature regime, limiting the ability of circuit designers to optimize cryo-CMOS circuits. Previous works have proposed deep-cryogenic models for MOSFETs\cite{beckers2018cryogenic, jazaeri2019review, beckers2020physical} in regimes of operation where the devices exhibit a behavior qualitatively similar to the conventional temperature regimes. However, MOSFETs operating at deep-cryogenic temperatures also exhibit novel behaviors, such as intersubband scattering \cite{casse2020evidence}, and drain current oscillations due to tunneling effects \cite{tripathi2022characterization}, \cite{han2021cryogenic}.

In this work, we present a deep-cryogenic modeling methodology for the GlobalFoundries 22-nm FDSOI technology based on a modified version of the BSIM-IMG compact model \cite{noauthor_bsim-img_2025}. The work involves modification of the compact model code and the introduction of an optimized model parameter extraction flow, for the cryogenic regime. The methodology for extracting static parameters via DC measurements, including those related to intersubband scattering, is discussed in Section \ref{DC}. The methodology for extracting dynamic parameters via RF measurements is discussed in Section \ref{RF}. Finally, in Section \ref{Simulation}, we use the custom model and extracted model cards to simulate a current-output digital-to-analog converter (DAC) and compare the simulated results versus measurement results.

\section{\textbf{DC Modeling Flow}} \label{DC}
\subsection{DC Measurements}\label{DC Meas}
To characterize the DC response at deep-cryogenic temperatures, we used large-scale test arrays with a distributed multiplexing architecture. Each test array contains 1024 cells, with each cell containing a MOSFET with varying geometric parameter values (width, length, and number of fingers), and multiplexing circuitry \cite{eastoe2024method}. In the test array there are several cells of each MOSFET parametric variant. This approach is compatible with dilution refrigerator setups, enabling data collection down to milli-Kelvin ambient temperature.  

For each parametric variant, the median current value is used to represent the typical I-V response. The model parameters for all MOSFETs were extracted under isothermal conditions, that is, at a fixed ambient temperature, with the temperature-scaling parameters disabled. To avoid transient heating effects after high-dissipation bias points, a cooling period was introduced between successive measurement points. 

The substrate of the die containing the test arrays was glued to a large PCB pad with conducting silver epoxy while individual signal pads were wire-bonded to PCB metal tracks connecting to cabling to room-temperature electronics. The PCB was placed at the mixing chamber stage of a dilution fridge with an ambient temperature of $\sim$15~mK. A Keithley 2636B SMU is used for voltage biasing and current measurement.

\subsection{Static Parameter Extraction} \label{Static}
Starting from the BSIM-IMG standard parameter value extraction flow \cite{bsimimg_extract}, we developed a simplified deep-cryogenic static parameter extraction flow.

\begin{figure}
\begin{minipage}{.5\textwidth}
   \centering
    \includegraphics[width=1.0\linewidth]{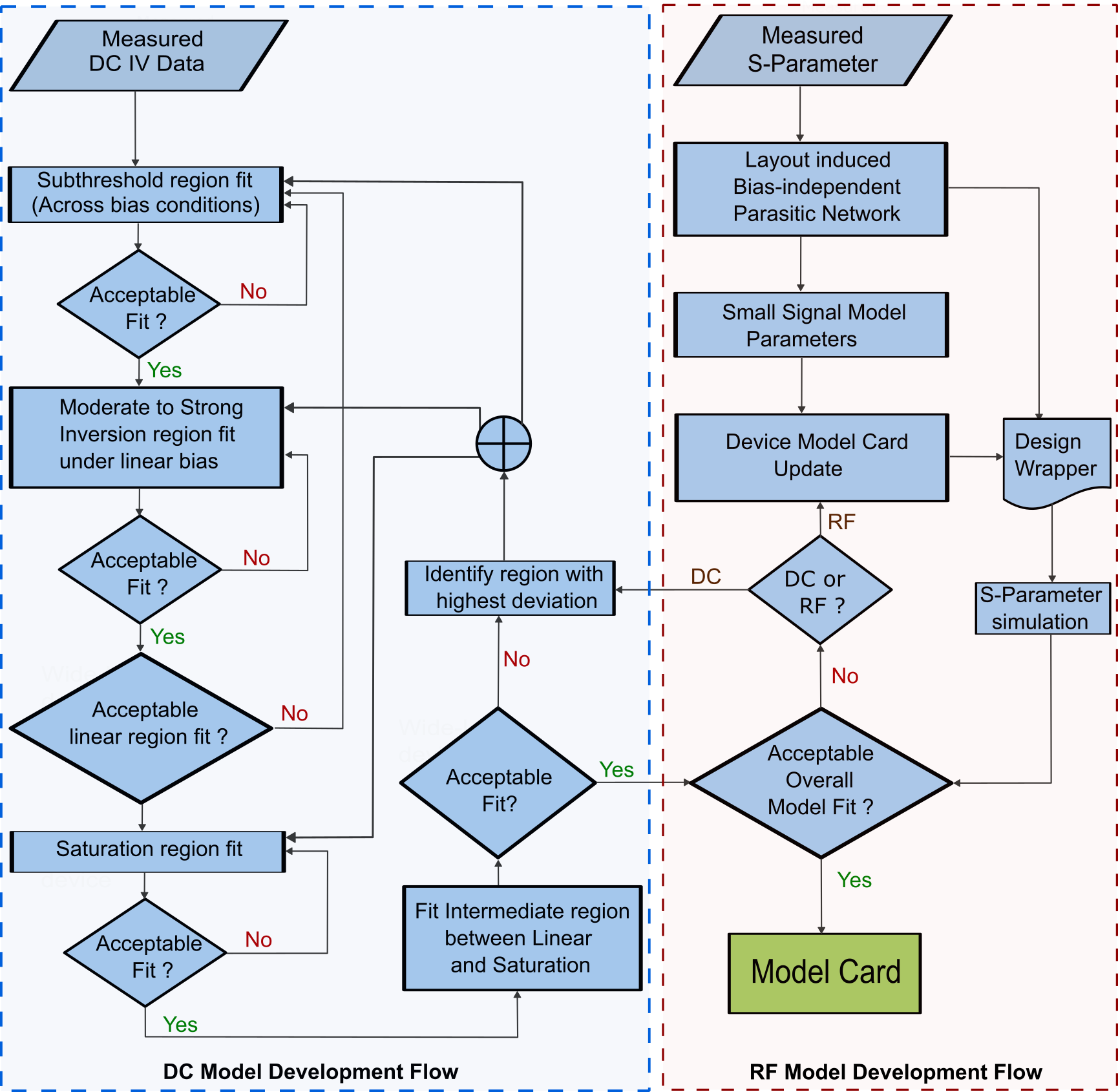}
    \caption{{MOSFET model card parameter extraction flow incorporating DC (left) and RF (right) sub-flows.} }
    \label{fig:modelflow} 
\end{minipage}
\end{figure}

The extraction flow, which is summarized in the left part of Fig. \ref{fig:modelflow}, is described below. First, the threshold voltage and subthreshold slope parameters are extracted through subthreshold-region fitting across multiple drain and back-gate biases. Secondly, the low-field mobility and mobility degradation parameters are adjusted using the drain current in moderate and strong inversion in the linear region ($V_{\mathrm{ds}}=50$ mV) for different back-gate voltages. Thirdly, the velocity saturation model parameters are adjusted using the drain current response in the saturation region. Finally, the smoothing parameters are adjusted to achieve good fit in the transition region between the linear and saturation regions. The extraction flow described above is repeated sequentially multiple times using the parameter values of the previous iteration as a starting point until a good-quality fit is achieved. A maximum error threshold, in terms of absolute, RMS, or percentage error, can be used to determine the quality of a model fit. While the goal is to minimize error, the error threshold should be set reasonably to prevent iterations that do not produce significant improvements.

In addition to the flow described above, the dimension-scaling parameters are obtained through the following sequential procedure. First, the extraction flow is run on a wide-long device. This reduces the number of parameters to be fit, as narrow-width and short-channel effects can be neglected. Subsequently, the short-channel model parameters are extracted, beginning with the linear subthreshold region, followed by the linear inversion region, and concluding with the saturation region. Geometry-based binning was required for certain parameters to achieve an accurate model fit across all geometries. 

\begin{figure}
    \begin{minipage}{.5\textwidth}
        \centering
        \includegraphics[width=1.0\linewidth]{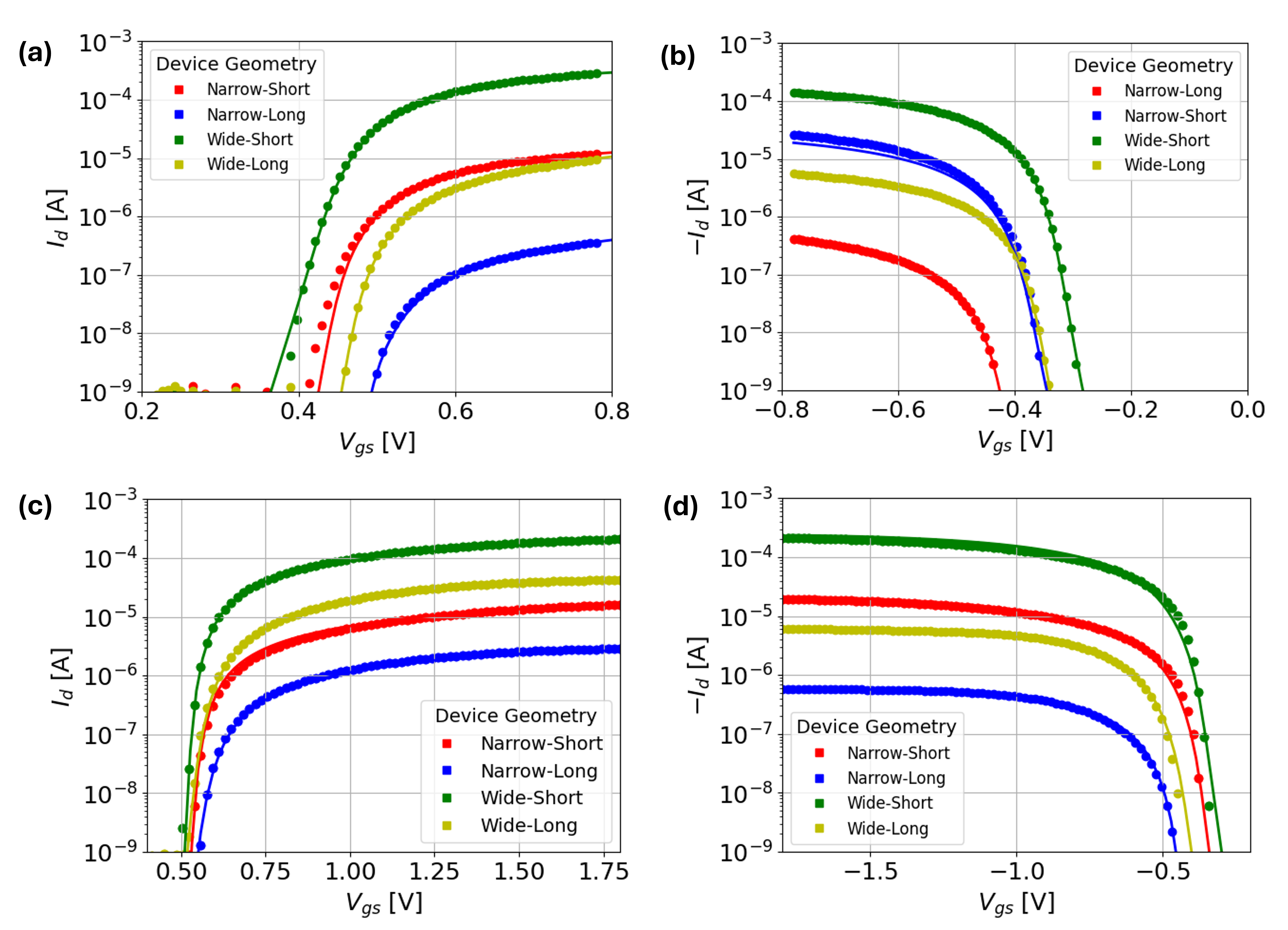}
        \caption{{
        DC $I_\mathrm{d}-V_\mathrm{gs}$ sweeps at $\sim$15~mK ambient temperature in the linear region ($V_\textrm{ds}=50$ mV) for four corner device geometries of: \textbf{(a)} thin-oxide n-type MOSFETs; \textbf{(b)} thin-oxide p-type MOSFETs; \textbf{(c)} thick-oxide n-type MOSFETs; and \textbf{(d)} thick-oxide p-type MOSFETs. Solid lines and circles correspond to typical simulated and measured drain current values, respectively.
        }}
        \label{fig:dcmodelresults_crnrgeo}    
    \end{minipage}       
\end{figure}

\begin{figure}
        \centering
        \includegraphics[width=1.0\linewidth]{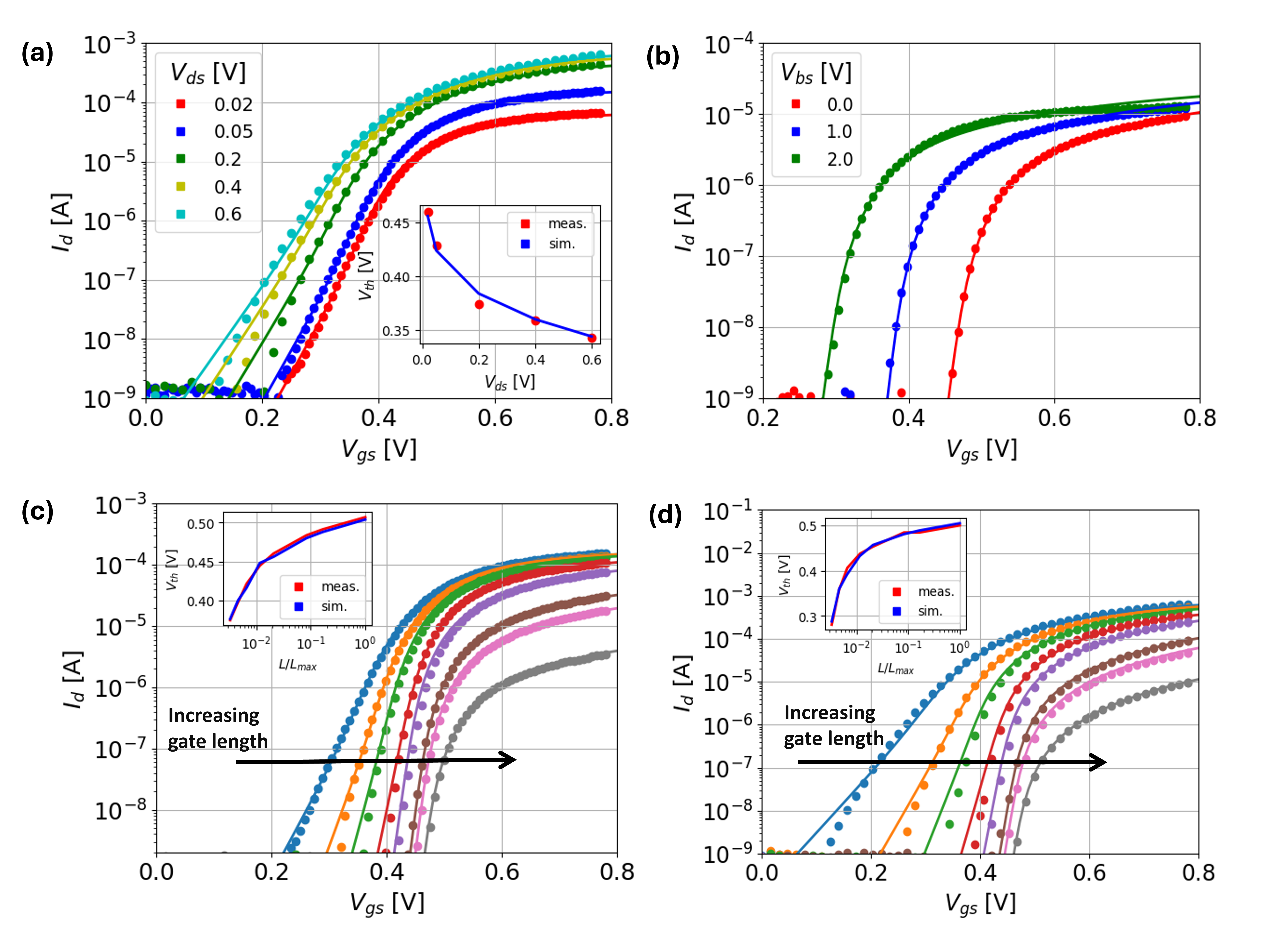}
        \caption{{DC $I_\mathrm{d}-V_\mathrm{gs}$ sweeps at $\sim$15~mK ambient temperature for thin-oxide n-type MOSFETs. Solid lines and circles correspond to typical simulated and measured drain current values, respectively. \textbf{(a)} Effect of drain bias for a short/moderate-width device geometry. \textbf{(b)} Effect of back-gate bias for a wide/long device geometry in the linear region ($V_\mathrm{ds}=50$ mV). \textbf{(c,d)} Effect of channel length for constant-width (wide) device geometries in the \textbf{(c)} linear region ($V_\mathrm{ds} = 50$ mV); and \textbf{(d)} saturation region ($V_{ds} = 800$ mV). Insets: $V_\mathrm{th}$ vs. normalized channel length.
        }}
        \label{fig:dcmodelresults}    
\end{figure}

To validate the fit obtained with the custom cryogenic modeling flow, we show a comparison between measured and simulated results using the extracted parameters at several bias conditions in Fig. \ref{fig:dcmodelresults_crnrgeo} and  Fig. \ref{fig:dcmodelresults}.

\begin{figure}
        \centering
        \includegraphics[width=0.8\linewidth]{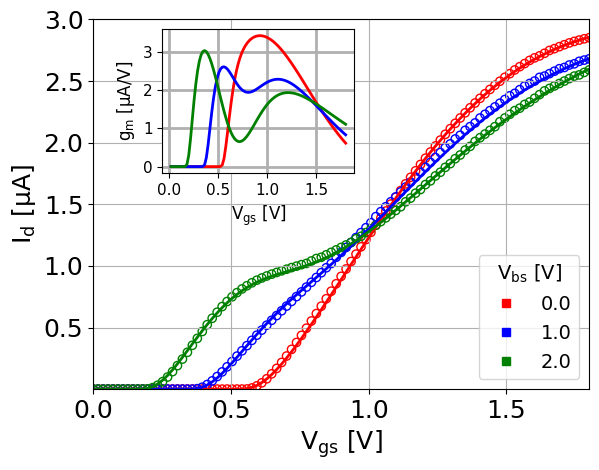}
        \caption{{Linear region $I_\mathrm{d}-V_\mathrm{gs}$ sweeps of a thick-oxide n-type narrow/long device geometry under different back-gate bias conditions. Mobility degradation due to intersubband scattering can be observed for higher back-gate bias voltage. The inset plot shows the corresponding calculated DC trans-conductance ($g_\mathrm{m}$) vs. $V_\mathrm{gs}$.
        }}
        \label{fig:dcmodelresults_eg}    
\end{figure}

\subsection{Intersubband Scattering Modeling} \label{Intersubband}
\begin{figure}
    \begin{minipage}{.5\textwidth}
        \centering
        \includegraphics[width=1.0\linewidth]{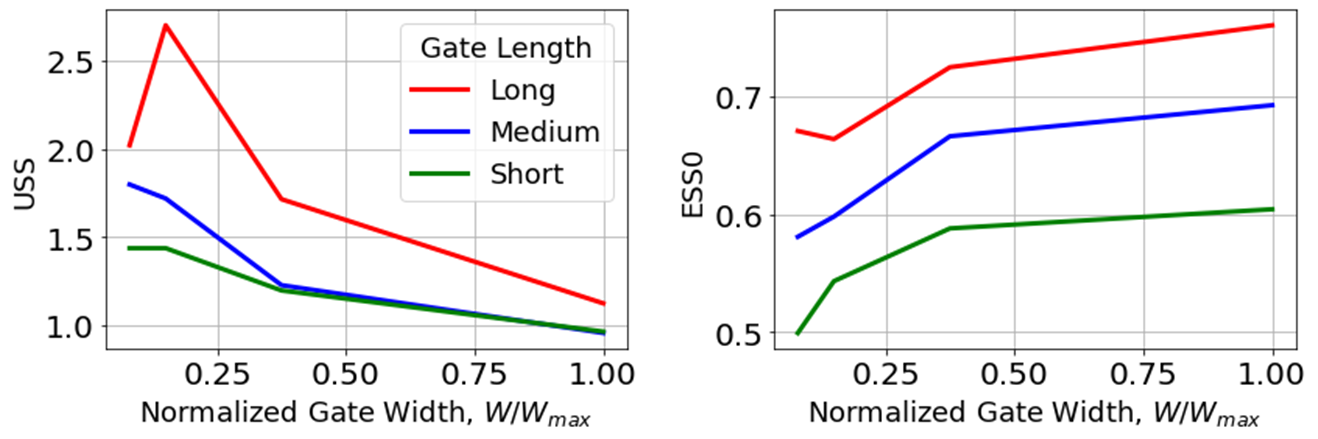}
        \caption{{Extracted $USS$ and $ESS0$ parameter values for different geometries of thick-oxide n-type MOSFETs.}}
        \label{fig:ussextrct}
    \end{minipage}
\end{figure}

At deep-cryogenic temperature, mobility degradation due to intersubband scattering \cite{casse2020evidence} is observed for thick-oxide devices operating in the linear region with a high back-gate bias as shown in Fig \ref{fig:dcmodelresults_eg}f. We enable accurate modeling of this phenomenon by introducing an additional term $D_{mobss}$ in the mobility degradation equations,
\begin{equation}    
        USS_{i} = USS \left( \frac{\left(V_{bgxpos}\right)^{USSB}}{1 + |E_{effm}|^{EU}} \right)  \left( V_{dsx} + 0.001 \right)^{USSD} \label{eq:dmobss}    
\end{equation}
\begin{equation}
        D_{mobss} = \frac{USS_{i}}{USS0 + \exp{\left(-\Delta ESS \left(V_{fgs} - ESS0\right)\right)}} \label{eq:dmobss2}    
\end{equation}
where $USS$, $ESS0$, $\Delta ESS$, $USSB$, $USSD$, $E_{effm}$, $EU$ and $USS0$ are additional parameters and $V_{fgs}$, $V_{bgxpos}$, and $V_{dsx}$ are the standard BSIM-IMG preconditioned terminal voltages for the front gate, back gate, and drain terminals, respectively. $USS$ models the relative magnitude of the intersubband scattering effect; $ESS0$ and $\Delta ESS$ model the front-gate bias dependency; and $USSB$ and $USSD$ model the back-gate and drain bias dependencies, respectively. The effective electric field, $E_{effm}$, and the parameter, $EU$, model the increase in mobility as $V_{fg}$ approaches the strong inversion region. $USS0$ is used to improve model fit, and an offset is introduced to $V_{dsx}$ to avoid discontinuities. Fig. \ref{fig:ussextrct} shows the extracted values of $USS$ and $ESS0$ for different device geometries. Larger values of $USS$ are extracted for longer devices, indicating an increase in the scattering effect as reported in prior studies \cite{aouad:tel-03852447}. We also observe an increase in the scattering effect with decreasing channel width.

\subsection{Parameter Variability Extraction} \label{Variability}

Accurate modeling of device parameter variability at deep-cryogenic temperatures is crucial, since it exhibits significant degradation in this temperature regime \cite{t_hart_characterization_2018, mizutani_measurement_2025}.
For this purpose, the test arrays include several instances of each geometric variant. The instances are laid out within the array such that the common centroid of each set of instances is at the geometric center of the array \cite{eastoe2024method}.

\begin{figure}   
    \begin{minipage}{.5\textwidth}
        \centering
        \includegraphics[width=1.0\linewidth]{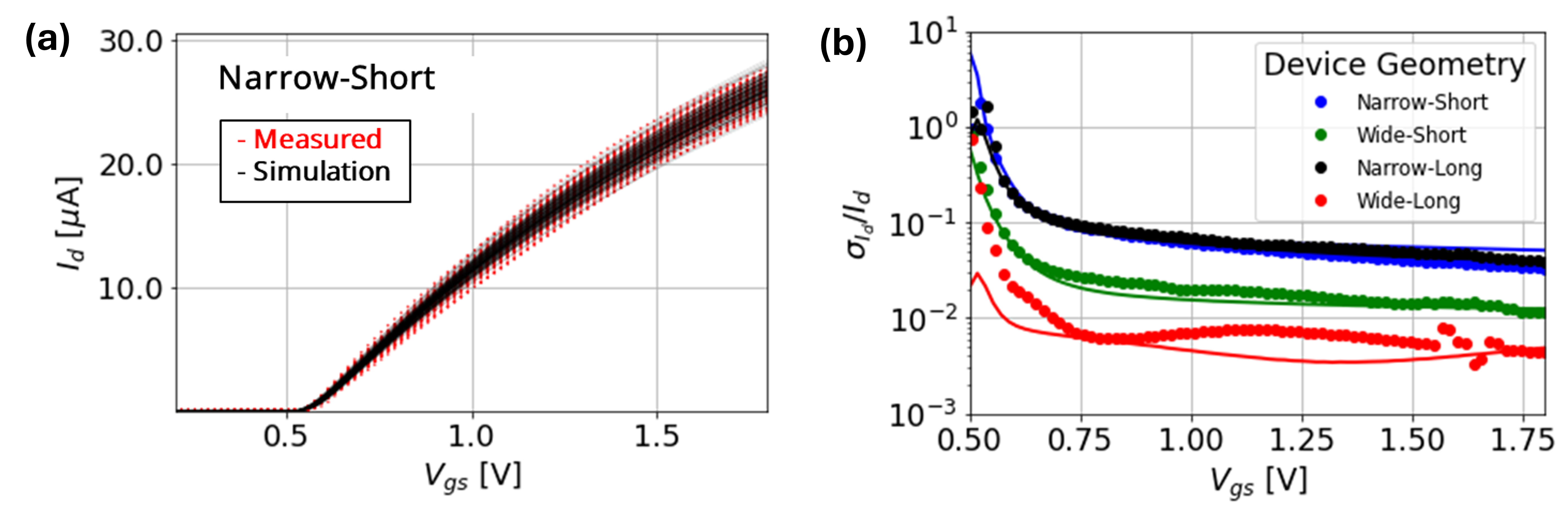}
        \caption{{Variation of static drain current response in linear region for thick-oxide n-type MOSFET devices. $I_\mathrm{d}$-$V_\mathrm{gs}$ sweeps from Monte Carlo mismatch simulation are in good agreement with measured data (from a single die) across device dimensions.} }
        \label{fig:device_var}
    \end{minipage}
\end{figure}    

 The measurements from a test array are able to capture intra-die variation, which is a subset of global process variability \cite{840854}. The standard deviation of the parameters is determined assuming that the distribution of each parameter is independent and imposing the constraint that the Monte Carlo simulations of the device I-V characteristics for a given geometry closely agree with the measured variability as shown in Fig. \ref{fig:device_var}.

\section{\textbf{RF Modeling Flow}} \label{RF}
\subsection{RF Measurements} \label{RF Meas}
The deep-cryogenic dynamic response of the MOSFETs is characterized using on-wafer probing. The MOSFETs are embedded in two-port RF Ground-Signal-Ground test structures, with the front gate connected to port 1, the drain connected to port 2, the source connected to the RF ground, and the back gate connected to a DC bias line. Additionally, on-wafer open and short calibration structures were included to move the measurement reference plane to the bottom metal connecting to the device terminals.

A Lakeshore TTPX cryogenic probe station was used to probe the structures, achieving an approximate ambient temperature of 4~K. Although the employed system is not capable of achieving DUT temperatures in the milli-Kelvin regime, the MOSFET behavior is not expected to change significantly between 4~K and milli-Kelvin temperatures \cite{cassedeliverable}. RF measurements in the milli-Kelvin regime are the subject of ongoing research, with solutions being explored based on on-wafer RF probing \cite{wei2023, 10.1063/5.0139825} or based on embedding the DUT in a setup with cryogenic switches \cite{9762297}.

For the RF measurements, the substrate of the die was glued to a large PCB pad with conducting silver epoxy. The back gate DC bias was applied through a wire-bonded connection. Wide-band bias-tees were used to combine DC biases with the RF signals before connecting to the probe.



\begin{figure}
    \begin{minipage}{.5\textwidth}
        \centering
        \includegraphics[width=0.8\linewidth]{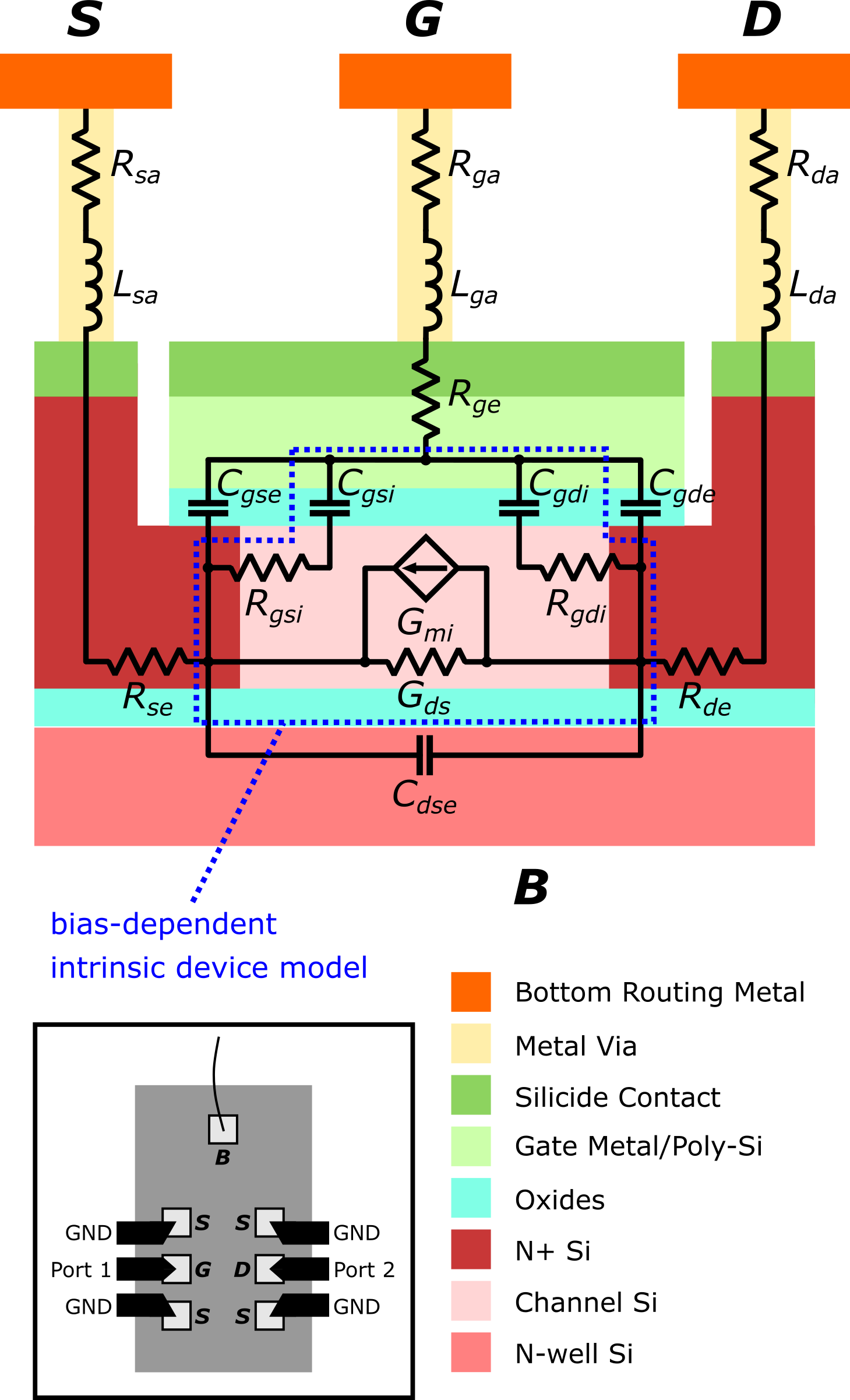}
        \caption{{Simplified RF small-signal equivalent circuit model of flip-well N-type FDSOI MOSFET overlaid on the device cross-section diagram (not to scale). Inset: representative top-down view (not to scale) of probing setup.}}
        \label{fig:ssem}
    \end{minipage}

\end{figure}

\begin{figure}
        \centering
        \includegraphics[width=1.0\linewidth]{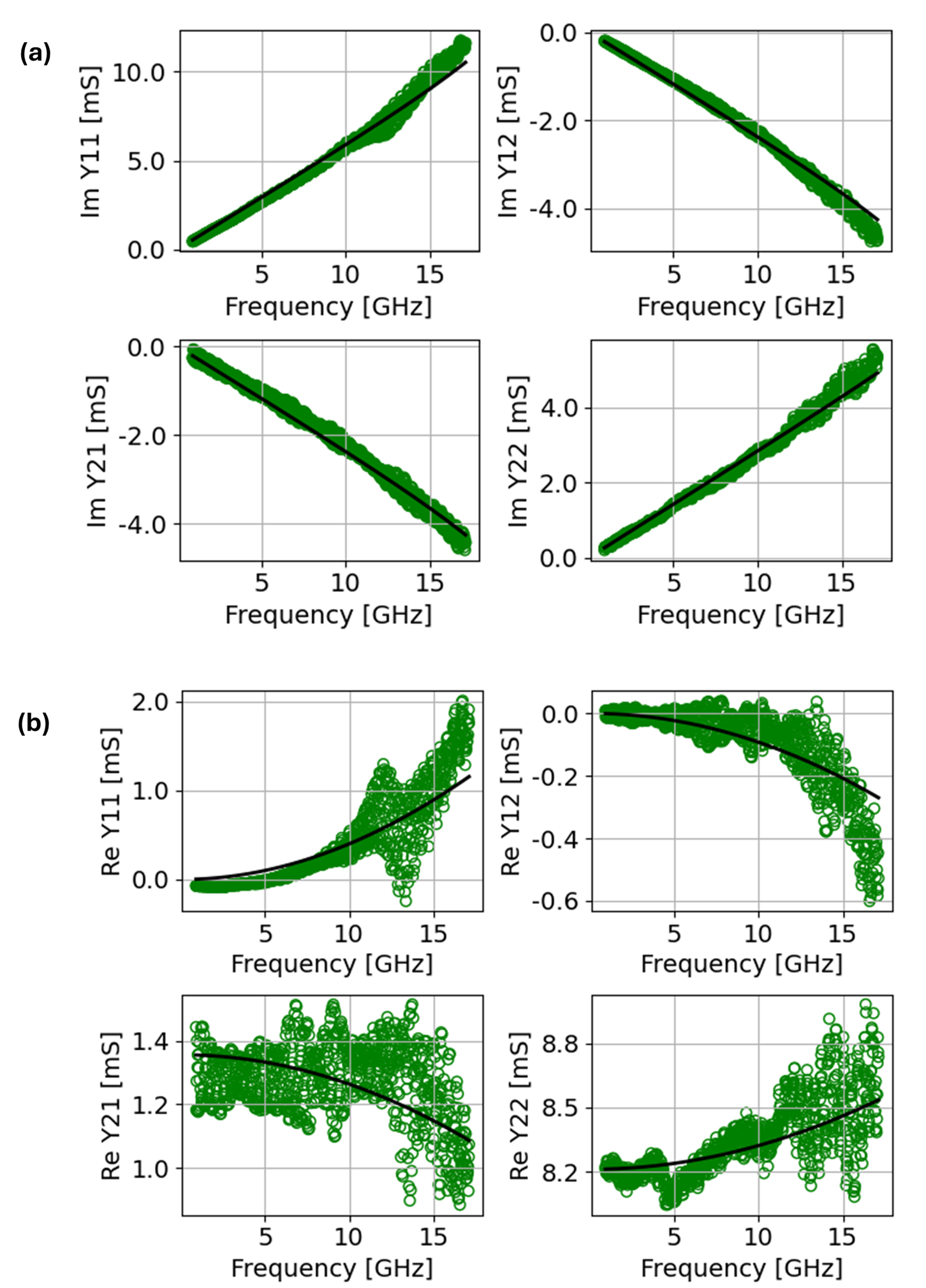}
        \caption{{Measured (green circles) and model fit (solid black lines) Y-parameters for a multi-finger thin-oxide n-type MOSFET biased at $V_\mathrm{gs}=0.8$ V, $V_\mathrm{ds}=0.8$ V, and $V_\mathrm{bs}=0$ V. Y-parameters' \textbf{(a)} real parts; and \textbf{(b)} imaginary parts. }}
        \label{fig:rfmodelfit}
\end{figure} 

\subsection{RF Parameter Extraction}\label{RF Pars}
In this work, we follow the RF small-signal modeling method described in \cite{wchakraborty22nmfdsoi}. The bias-independent extrinsic parasitic elements are determined first from the Y-parameters under zero bias ($V_\mathrm{gs}=0$~V, $V_\mathrm{ds}=0$~V). Subsequently, the intrinsic bias-dependent parameters are determined using Y-parameter response under different active bias conditions. Generating a simplified set of analytically-solvable Y-parameter expressions corresponding to the RF small-signal equivalent circuit shown in Fig. \ref{fig:ssem} is tedious to achieve. Instead, a heuristic-based approach using the PyTorch Adam optimizer was used to extract the small-signal model parameters effectively. The resulting model fits for a thin-oxide n-type MOSFET are shown in Fig. \ref{fig:rfmodelfit}.

\begin{figure}
    \begin{minipage}{.5\textwidth}
        \centering
        \includegraphics[width=0.8\linewidth]{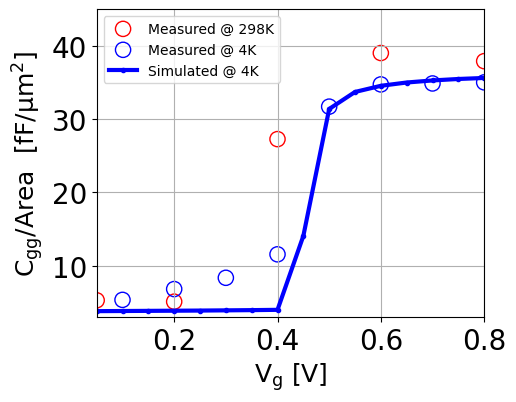}
        \caption{{ Front gate capacitance per unit area vs. bias voltage, for $V_\mathrm{ds}=0$ V and $V_\mathrm{bs}=0$ V, extracted from the RF measurement (circles) and simulated (solid line).
        }}
        \label{fig:cggvsvg}
    \end{minipage}

\end{figure}

The extracted RF small-signal model parameters are used to adjust the parameters in the BSIM-IMG model to improve device dynamic simulation accuracy. Parameters in the BSIM-IMG model that capture various overlap and fringe capacitance between device terminals are extracted utilizing the various capacitance values obtained from small-signal model fitting. As an example of this, Fig. \ref{fig:cggvsvg} shows the front gate capacitance under a bias condition of $V_\mathrm{ds}=0$ V and 4~K ambient temperature as extracted from RF measurements as $C_\mathrm{gg}\approx\mathfrak{Im}[Y_{11}]/\omega$ and compared with the simulated results using the extracted BSIM-IMG model. The simulated gate capacitance per unit area is in good agreement with measured data. Minimal change is observed between room temperature and 4~K at low-bias and high-bias conditions, in agreement with prior studies \cite{seidel202322fdx}. Similarly, parameters modeling resistance introduced by the source and drain regions are also adjusted. Note that these steps might require re-adjusting the mobility model parameters (extracted previously using the DC modeling sub-flow). A bias-independent parasitic network can be introduced in the design wrapper to model the parasitic effects outside the scope of the BSIM-IMG model. The integration of the RF sub-flow into the full modeling flow is shown in Fig. \ref{fig:modelflow}.

\section{\textbf{Simulation Demonstration}} \label{Simulation}
\begin{figure}
    \begin{minipage}{.5\textwidth}
        \centering
        \includegraphics[width=1\linewidth]{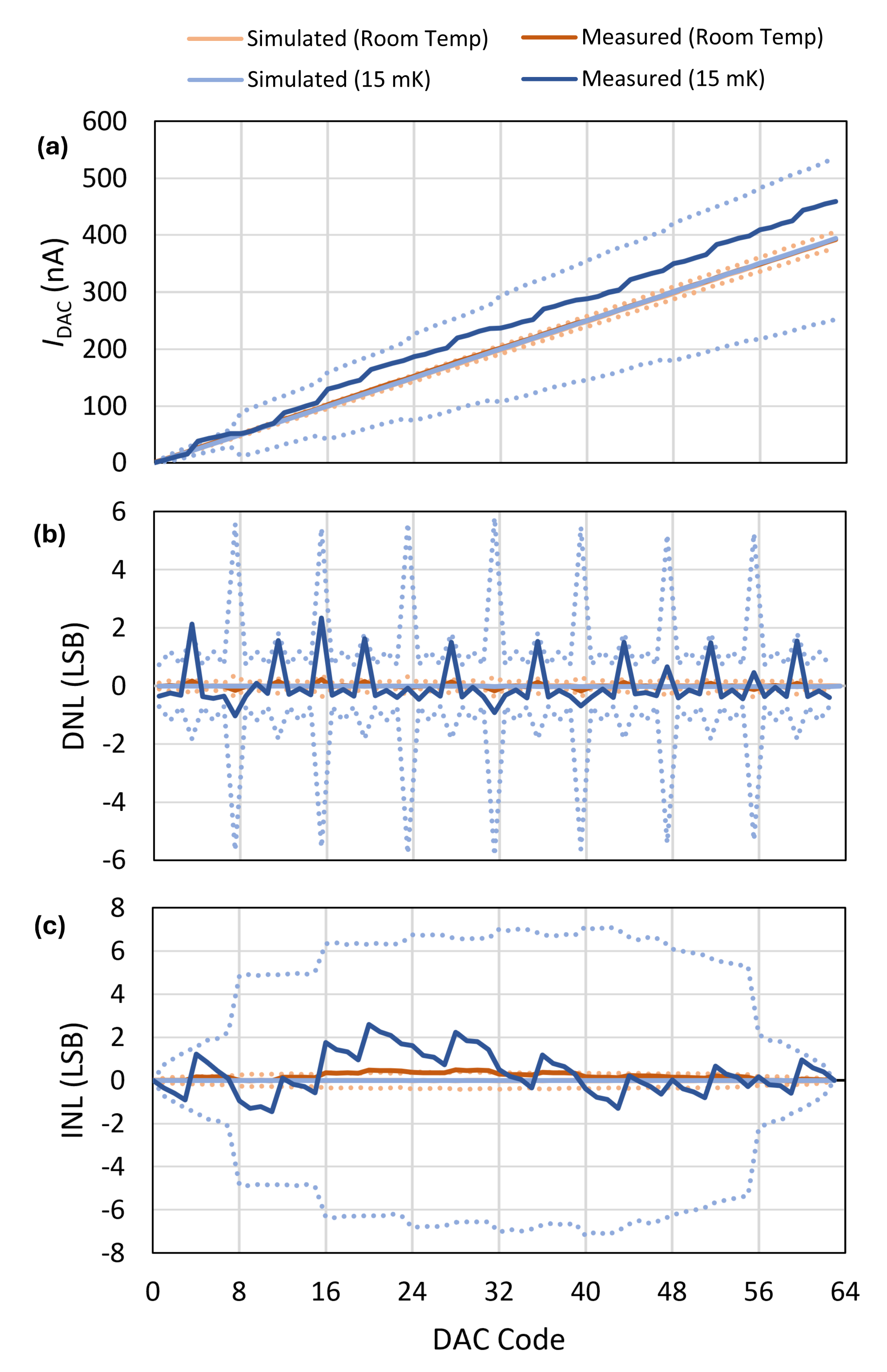}
        \caption{DAC simulated and measured results at 293~K and 15~mK ambient temperature. For simulated data, the solid lines represent the typical device behavior while the dotted lines represent the $\pm$3$\sigma$ bounds from the Monte Carlo simulation. \textbf{(a)} Transfer characteristic; \textbf{(b)} differential non-linearity; and \textbf{(c)} integral non-linearity.}
        \label{fig:tsns_sims}
    \end{minipage}

\end{figure}

To demonstrate the utility of the developed deep-cryogenic MOSFET models, the simulation results of a current-output DAC circuit \cite{olivieri2025integrated} obtained using the extracted deep-cryogenic models were compared with the 15~mK measurement data of one DAC part. The comparison is shown in Fig. \ref{fig:tsns_sims}. While using typical-only models for this simulation results in near-perfect linearity, significant non-linearity is observed in the measurement and the Monte-Carlo-enabled simulations. The measured data is within the statistical 3-sigma bounds of the simulated results. For a higher statistical confidence on the accuracy of the extracted variability parameters, measurements of additional parts would be required. To the authors’ knowledge, this is the first work to numerically quantify the impact of device variability at deep-cryogenic temperatures in the performance of an integrated circuit.

\section{\textbf{Conclusion}} \label{Conclusion}
In this work, we have introduced a custom deep-cryogenic modeling flow for MOSFETs based on BSIM-IMG and applied it to the parameter extraction of a 22-nm FDSOI technology. The presented flow involves DC and RF modeling sub-flows which enable the extraction of static and dynamic model parameters, respectively. By modifying the BSIM-IMG mobility equations, we are able to capture intersubband scattering, a deep-cryogenic effect.

We identify several limitations of the presented work. The used isothermal approach does not enable the modeling of thermodynamic phenomena such as device self-heating or cross-device heating, and would require independent extraction of model parameters at each ambient temperature. 

We also identify paths for expansion of the presented work, such as characterization of noise, layout-dependent effects, passives and back-end-of-line parasitics. Future model enhancements will also be required to address other deep-cryogenic effects, such as subthreshold drain current oscillations.

In conclusion, the presented deep-cryogenic modeling capability represents a significant improvement in the prediction capabilities of the cryo-CMOS circuit design flow, increasing the potential for first-time-right silicon. 

\section*{Acknowledgment}
The authors acknowledge Chris Gamlath, Charan Kocherlakota, James Kirkman, Jonathan Warren, and Alexander Waterworth of Quantum Motion for their technical support. 

\section*{Author Contributions}

D. Dutta acquired the data. All authors analyzed the data. D. Dutta packaged the dies. A. Gomez-Saiz and K. Ture conceived and designed the DC and RF test structures. D. Dutta developed the model mobility modifications. F. Olivieri designed and simulated the cryogenic DAC circuit. A. Gomez-Saiz and G. M. Noah supervised the work. All authors contributed to the writing of this article.

\bibliography{reference}

\end{document}